\documentclass{iaus}
\usepackage{graphicx}

\title[FSS Galaxies: Supernova Ia Rates] 
{Five Supernova Survey Galaxies in the Southern Hemisphere: Supernova Ia Rates}

\author[A.~A.~Hakobyan et al.]   
{A.~A.~Hakobyan$^{1,2}$, A.~R.~Petrosian$^1$, G.~A.~Mamon$^3$, B.~McLean$^4$,
 \ D.~Kunth$^3$, M.~Turatto$^5$, E.~Cappellaro$^6$, F.~Mannucci$^7$, R.~J.~Allen$^4$,
 \ N.~Panagia$^{4,8,9}$, M.~Della~Valle$^{10,11}$, and G.~V.~Petrosyan$^2$}

\affiliation{$^1$Byurakan Astrophysical Observatory, 0213 Byurakan, Aragatsotn Province, Armenia
\\ email: {\tt hakobyan@bao.sci.am} \\[\affilskip]
$^2$Department of General Physics and Astrophysics, Yerevan State University,
\\1 Alex Manoogian, 0025 Yerevan, Armenia \\[\affilskip]
$^3$Institut d'Astrophysique de Paris (UMR 7095: CNRS \& UPMC), \\98 bis Bd Arago, 75014 Paris, France \\[\affilskip]
$^4$Space Telescope Science Institute, 3700 San Martin Drive, Baltimore, MD 21218, USA \\[\affilskip]
$^5$INAF-Osservatorio Astronomico di Trieste, Via Tiepolo 11, 34143 Trieste, Italy \\[\affilskip]
$^6$INAF-Osservatorio Astronomico di Padova, Vicolo dell'Osservatorio 5, 35122 Padova, Italy \\[\affilskip]
$^7$INAF-Osservatorio Astrofisico di Arcetri, Largo E.~Fermi 5, 50125 Firenze, Italy \\[\affilskip]
$^8$INAF-Osservatorio Astrofisico di Catania, Via Santa Sofia 78, 95123 Catania, Italy \\[\affilskip]
$^9$Supernova Ltd., OYV \#131, Northsound Road, Virgin Gorda, British Virgin Islands \\[\affilskip]
$^{10}$INAF-Osservatorio Astronomico di Capodimonte, Salita Moiariello 16, 80131 Napoli, Italy \\[\affilskip]
$^{11}$International Center for Relativistic Astrophysics, Piazzale della Repubblica 2,\\ 65122 Pescara, Italy}

\pubyear{2011}
\volume{281}  
\jname{Binary Paths to Type Ia Supernovae Explosions}
\editors{R.~Di~Stefano, M.~Orio, \& M.~Moe, eds.}
\begin{document}

\maketitle

\begin{abstract}
Based on the database with 56 supernovae (SNe) events discovered in 3838 galaxies of the southern hemisphere,
we compute the rate of SNe of different types along the Hubble sequence normalized
to the optical and near-infrared (NIR) luminosities as well as to the stellar mass of the galaxies.
We find that the rates of Type Ia SNe show a dependence on both morphology and colors of the galaxies,
and therefore, on the star-formation activity. The rate of SNe~Ia can be explained by assuming
that at least 15\% of Ia events in spiral galaxies originate in relatively young stellar populations.
We also find that the rates show no modulation with nuclear activity or environment.
\end{abstract}


Using the database
of the Five SN Survey (FSS) galaxies identified in the field of
Deep Near Infrared Survey (DENIS) of the southern sky
(\cite[Hakobyan~\emph{et~al.}~2009]{2009Ap.....52...40H}),
we compute the rates of SNe Ia by grouping galaxies into four classes:
E-S0, S0/a-Sb, Sbc-Sd, and Sm-Irr, and normalizing rates
to the optical \emph{U}, \emph{B}, \emph{R}, and NIR
\emph{I}, \emph{J}, \emph{H}, \emph{K} luminosities as well as
to the stellar mass of the galaxies (\cite[Hakobyan~\emph{et~al.}~2011]{2011arXiv1104.0300H}).
We compute also the rate of SNe for
the samples of the galaxies with different level of nuclear activity and
with different properties of their local environment
(\cite[Hakobyan~\emph{et~al.}~2011]{2011arXiv1104.0300H}).
Close inspection of the results reveals the following:

1) The rate of SNe Ia per unit mass increases by a factor of about 2.8
from E-S0 to\, Sbc-Sd galaxies. A similar trend can be seen when the galaxies are
binned according to their $U-B$, $B-R$, and $B-K$ colors.
Within the uncertainties, the SNe~Ia rate is independent on the host galaxy
$R-I$, $I-J$, $J-H$, and $H-K$ colors (see also
\cite[Cappellaro~\emph{et~al.}~1999]{1999A&A...351..459C}).
In $B-K$, the ratio between SN~Ia
rates in galaxies bluer than $B-K = 2.8$ and redder than $B-K = 4.2$ is larger than 20,
which is comparable with the values of $\sim30$ times of
\cite[Mannucci~\emph{et~al.}~(2005)]{2005A&A...433..807M} and $\sim15$ times of
\cite[Li~\emph{et~al.}~(2011)]{2011MNRAS.412.1473L}.
The existence of such differences in SNe Ia rates between late spirals and
ellipticals implies that the frequency of progenitors exploding as a SNe Ia
per unit time changes considerably with the ages of the parent population of the galaxies.

2) In \cite[Mannucci~\emph{et~al}.~(2005)]{2005A&A...433..807M},
a simple toy model was introduced in which the rate of Type Ia~SN is
reproduced adding a constant contribution from ``old'' progenitors, independent of color
and fixed at the value measured in the ellipticals, plus a contribution proportional to
the rate of core-collapse (CC) SNe. The best fitting agreement between observed ${\rm SNuM} = f(B-K)$
$({\rm SNuM} \equiv 10^{-12} \,\, {\rm SNe \,\, M^{-1}_{\odot} \,\, yr^{-1}})$
and the toy model curves is obtained for the ``young'' progenitors fraction value of $(35\pm8)\%$
(\cite[Mannucci~\emph{et~al.}~2005]{2005A&A...433..807M}).
A similar analysis (\cite[Li~\emph{et~al.}~2011]{2011MNRAS.412.1473L})
produces a smaller (but statistically consistent) ``young''
progenitors fraction of $(22\pm7)\%$. We have performed the same estimation using a slightly
different approach. We assume that the ``old'' SNe~Ia progenitors belong to the bulge population
of the galaxies and we fix the rate of SN~Ia in bulges to that in galaxies with
red $B-K$ colors (0.01~SNuM). We also assume that the ``young'' progenitors of SN~Ia belong
to the disk population of the galaxies.
Taking the recent determination (\cite[Oohama~\emph{et~al.}~2009]{2009ApJ...705..245O})
of bulge to total mass ratios (B/T) for different
types of galaxies, we estimated for the ``young'' SNe~Ia progenitors fraction a value
of $(15\pm7)\%$ for early-type spirals, and $(20\pm8)\%$ for late-type spirals.
This result is in agreement, within the errors, with the ``prompt'' fraction of
SNe~Ia previously reported.

3) We also computed the SNe rates in units of stellar mass after binning the galaxies
according to their activity level. We found that there is no significant
difference between the rate of SNe Ia in normal
$(0.10\pm0.04 {\rm \,\ SNuM})$ and in active or star-forming (A/SF)
$(0.08\pm0.05 {\rm \,\ SNuM})$ galaxies.

4) Using SNe as tracers of star-formation, we addressed also the problem of the
relation between galaxies interaction and star-formation. We
presented SNe rates, in units of stellar mass, for galaxies without $(n = 0)$, and with
at least one neighboring object $(n > 0)$. Comparing the rates of SNe in galaxies without
and with neighbor(s), we found that there is no significant difference between the rate
of SNe Ia in galaxies with $n = 0$ $(0.09\pm0.04 {\rm \,\ SNuM})$ and with $n > 0$
$(0.09\pm0.05 {\rm \,\ SNuM})$ neighbors.

\acknowledgments{The work by A.A.H. and A.R.P. was part of the
Collaborative Bilateral Research Project of the State Committee of Science (SCS)
of the Republic of Armenia and the French Centre National de la Recherch\'{e} Scientifique (CNRS).
This work was made possible in part by a research grant from the
Armenian National Science and Education Fund (ANSEF) based in New York, USA.
A.A.H. and G.V.P. wish to thank the Calouste Gulbenkian Foundation
for the travel grant to attend the IAU Symposium 281.}

\end{document}